\documentclass[letterpaper]{IEEEtran}

\usepackage[english]{babel}
\usepackage{amsmath,amssymb}
\usepackage{times}
\usepackage{relsize}
\usepackage{graphicx}
\usepackage{url}
\usepackage{booktabs}
\usepackage{multirow}
\usepackage{mparhack}
\usepackage{subfigure}
\usepackage{authblk}
\usepackage{amsthm}

\usepackage[noend]{algorithmic}
\usepackage[linesnumbered,ruled,vlined]{algorithm2e}

\usepackage{array}
\usepackage{cite}
\usepackage{eqparbox}
\usepackage{mdwmath}
\usepackage{balance}
\usepackage{epsfig}
\usepackage{xcolor}

\usepackage{mathtools}
\usepackage{bm}
\usepackage{amsfonts}
\usepackage[all]{xy}
\usepackage{etoolbox}
\usepackage{graphicx}
\DeclareMathAlphabet{\mathantt}{OT1}{antt}{li}{it}
\DeclareMathAlphabet{\mathpzc}{OT1}{pzc}{m}{it}

\usepackage{accents}

\DeclareFontFamily{OT1}{pzc}{}
\DeclareFontShape{OT1}{pzc}{m}{it}%
  {<-> s * [1.1] pzcmi7t}{}
\DeclareMathAlphabet{\mathpzc}{OT1}{pzc}%
                     {m}{it}

\newenvironment{sequation}{\begin{equation}\small}{\end{equation}}

\DeclareMathOperator{\argmin}{\arg\min}

\makeatletter
\patchcmd{\@maketitle}
  {\addvspace{0.5\baselineskip}\egroup}
  {\addvspace{-1.45\baselineskip}\egroup}
  {}
  {}
\makeatother

\title{Device Caching for Network Offloading: Delay Minimization with Presence of User Mobility}
\author{
Tao~Deng,
Lei~You,
Pingzhi~Fan,~\IEEEmembership{Fellow,~IEEE},
 and Di~Yuan,~\IEEEmembership{Senior Member,~IEEE}

\thanks{The work of T. Deng and P. Fan is supported by the National Science Foundation of China (NSFC, No. 61731017), the 111 Project (No. 111-2-14), and the China Scholarship Council (CSC). The work of L. You and D. Yuan is supported by the Swedish Research Council.}
 \thanks{T.\ Deng and P.\ Fan  are with the School of Information Science and Technology, Southwest Jiaotong University,
Chengdu, Sichuan 610031, China (e-mail: dengtaoswjtu@foxmail.com and p.fan@ieee.org). }%
\thanks{L.\ You and D.\ Yuan are with the Department of Information Technology, Uppsala University, 751 05 Uppsala, Sweden (e-mail:
\{lei.you, di.yuan\}@it.uu.se)}
}

\begin{document}

\maketitle

\begin{abstract}
A delay-optimal caching problem (DOCP) in device-to-device (D2D)
networks with moblity is modelled.  The
problem arises in the context of achieving offloading using device
caching, and the offloading effect is represented by the expected
network load ratio (NLR) which is the percentage of data that has to
be downloaded from the network.
Compared with the related studies, this work considers minimizing
delay with NLR guarantee in mobility scenarios.  A lower bound of
global optimum is derived, thus enabling performance benchmarking of
any sub-optimal algorithm.  For problem-solving, an effective search
algorithm (ESA) is proposed based on the bound. Simulations are
conducted to evaluate the effectiveness of the ESA algorithm.
\end{abstract}
\vspace{-1mm}

\IEEEpeerreviewmaketitle
\vskip -15pt
\vspace*{-0.1em}
\section{Introduction}

Recently, there is a considerable growth of the research interest in
caching. With caching, users can obtain the contents of interest from
edge devices \cite{XWang2014Cache}\cite{Fan2016Coping}.  The caching
performance depends heavily on the cache placement strategy.  In
mobility scenarios, it is necessary to consider the impact of mobility
on optimal caching design
\cite{RWang2016Mobility}.
In \cite{TWei2014MPCS}, the authors studied optimal caching placement
with respect to content popularity and user
mobility.  In \cite{YGuan2014Mo}, mobility-aware caching was studied
and an approximation algorithm was developed.  In
\cite{HLi2016Mobility}, based on mobility prediction, a seamless cache
handover framework was designed.  In \cite{KPoularakis2017}, the
authors studied the caching placement problem with mobility in
heterogeneous networks.  In \cite{RWang2016}, the authors considered
caching placement to maximize the data offloading ratio in
device-to-device (D2D) networks.  In \cite{TDeng2017CostJ}, the
authors considered a cost-optimal caching problem in D2D networks.  As
shown in \cite{RWang2016} and \cite{TDeng2017CostJ}, the offloading
effect benefits from allowing longer time for obtaining data between
devices. However, there is a clear trade-off between offloading and
delay. This letter aims to address this aspect.

In this letter, we model a delay-optimal caching problem (DOCP),
subject to an upper limit on the expected network load ratio (NLP), which is defined as the percentage of the data that has to be downloaded from the network.
 We prove that DOCP is equivalent to another optimization problem, referred to as the
  expected NLR-optimal caching problem (NOCP).
A lower-bounding function of the objective function of NOCP is derived, thus giving an approximative NOCP (ANOCP) with linearization.
The global optimum of ANOCP can be derived, resulting in a lower bound of the global optimum of DOCP.
More importantly, the lower bound enables performance benchmarking of any sub-optimal algorithm.
For problem-solving, an effective search algorithm (ESA) is proposed based on the lower bound.
Simulation results validate the effectiveness of the proposed algorithm.

\vspace{-1.5mm}
\section{System Model and Problem Formulation} \label{System_model}
\subsection{System Scenario and Caching Placement}
Let $\mathcal{U}=\{1,2,\dots,U \}$ denote the set of users, where each user $i$ has a local cache with size $C_i$.
When two users meet, they can communicate with each other, which is referred to as a contact. We assume that the contact between users $i$ and $j$ follows the Poisson distribution with rate $\lambda_{ij}$. This assumption is common \cite{RWang2016,TDeng2017CostJ,VConan2008}. Moreover, by investigating real-world mobility traces, the tail behavior of the inter-contact time distribution can be described by an exponential distribution \cite{Zhu2010Recognizing}.
Let $\mathcal{F}=\{1,2,\dots,F \}$ denote the set of files, where
each file $f$ is encoded into $S^f_\text{max}$ segments via a coding technique
\cite{KPoularakis2017,Leong2012Distributed}. File $f$, $f\in \mathcal{F}$, can be recovered by collecting at least $S^f_\text{rec}$ distinct segments \cite{RWang2016}.
The number of segments of file $f$ stored at user $i$ is denoted as
$x_{fi}$. The probability that user $i$ requests file $f$ is denoted
by $P_{fi}$, where $\sum_{f\in \mathcal{F}}P_{fi}=1$. When a user
requests file $f$, it collects the segments of $f$ from the
encountered users through D2D communications and from its own
cache. The former is subject to a delay period $T$.  At the end of
$T$, if the total number of collected segments of
this file is at least $S^f_\text{rec}$, this file can
be recovered. Otherwise, to reach $S^f_\text{rec}$ segments, the user
will have to download additional segments from the network.
Typically, compared to the waiting time to meet other users, the data
downloading time from the network is negligible.

When two users meet, each of them can collect up to $B$ segments from the other.
The number of contacts for users $i$ and $j$ is denoted by $M_{ij}$. Here, $M_{ij}$ follows a Poisson distribution with mean
$\lambda_{ij}T$.
Let $S_{fij}$ denote the number of segments of file
$f$ collected by user $i$ from user $j$ within $T$, which is $\min(BM_{ij},x_{fj})$.
Let $S_{fi}$ denote the number of segments of file $f$ collected by user $i$ from itself and all the other users within $T$, which is $\sum_{j\in \mathcal{U},j\neq i} \min(BM_{ij},x_{fj})+x_{fi}$.
If $S_{fi}<S^f_\text{rec}$, user $i$ has to download the remaining number of segments from the network.
Let $S^N_{fi}$ denote the number of segments of file $f$ downloaded from the network, which is
$\max(S^f_\text{rec}-S_{fi},0)$.
The NLR due to user $i$ requesting file $f$, denoted by $r_{fi}$, is
$S^N_{fi}/S^f_\text{rec}$.
The NLR that user $i$ recovers a file, denoted by $r_i$, is thus
$\sum_{f\in \mathcal{F}} r_{fi}P_{fi}$. Therefore,
the expected NLR to recover a file for all the users can be expressed as in (\ref{C}),
where $\mathbb{E}(\cdot)$ denotes the expectation operator.
\begin{sequation}
\begin{aligned}
R(\bm{x},T) \triangleq
 \mathbb{E}\{\frac{1}{U}\underset{i\in \mathcal{U}}\sum \underset{f\in \mathcal{F}}\sum P_{fi}[ \frac{\max(S^f_\text{rec}-S_{fi},0)}{S^f_\text{rec}} ] \}.
\label{C}
\end{aligned}
\end{sequation}

\subsection{Problem Formulation}
Our problem is to minimize $T$ via optimizing $\bm{x}$ for an upper limit of NLP, denoted by $R'$.
Thus, DOCP reads
\begin{figure}[!h]
\vskip -20pt
\begin{subequations}
\begin{alignat}{2}
\quad &
\min\limits_{\bm{x},T}\quad  T
 \label{F_e} \\
\text{s.t}. \quad
& R(\bm{x},T)  \le R', \label{F_a}\\
& {\sum_{f\in \mathcal{F}}} x_{fi}\le C_i, ~i\in \mathcal{U} \label{F_f}\\
& \underset{i\in \mathcal{U}}\sum x_{fi}\le S^f_\text{max}, ~f\in \mathcal{F} \label{F_h}\\
& x_{fi} = \{0,1,\dots,S^f_\text{rec}\}, ~i\in \mathcal{U}, ~ f\in \mathcal{F} \label{F_g}
\end{alignat}
\label{eq:mine}
\vskip -25pt
\end{subequations}
\end{figure}

Eq.~(\ref{F_a}) states that the expected NLR should respect the upper limit.
Eq.~(\ref{F_f}) ensures that the total number of cached segments does not exceed the cache capacity.
By Eq.~(\ref{F_h}), the total number of segments of a file, stored by all users, is bounded by the number of encoded segments, guaranteeing that the collected segments of any file will be distinct from each other.
Additionally, our problem is subject to a limit $T_\text{max}$ that is acceptable by the users.

\noindent \textbf{Lemma 1.} $[\bm{x}^*,T^*]$ is the optimum of DOCP only if $R(\bm{x}^*,T^*)=R'$.
\begin{proof} By the proof in Appendix, $R(\bm{x}^*,T)$ is a continuous and monotonically decreasing function with respect to $T$.
As a result, $R(\bm{x}^*,T^*)<R'$ is suboptimal, because with $x^*$ fixed, reducing $T^*$ is possible, and this improves the time objective without violating constraints (\ref{F_f}) and (\ref{F_h}).
\end{proof}

For fixed $T$, define an NOCP problem
\begin{equation}
\begin{aligned}
R^*(T)\triangleq \min\limits_{\bm{x}} R(\bm{x},T) |(\ref{F_f})-(\ref{F_g}),
\nonumber
\end{aligned}
\end{equation}
\noindent \textbf{Lemma 2.} For any $T'$, denote $\bm{x}'=\argmin_{\bm{x}} R(\bm{x},T')|(\ref{F_f})-(\ref{F_g})$. Then $[\bm{x}',T']$ is a
feasible solution of DOCP if and only if $R^{*}(T')\leq R'$.
\begin{proof}
For sufficiency, it is obvious that $[\bm{x}',T']$ satisfies all the constraints in DOCP.
This is because, $[\bm{x}',T']$ fulfills (\ref{F_a}) due to $R^{*}(T')\leq R'$. In addition, $[\bm{x}',T']$ is a feasible in NOCP
and therefore satisfies (\ref{F_f})-(\ref{F_g}).
For necessity, if $[\bm{x}',T']$ is a feasible in DOCP,
then it satisfies (\ref{F_f})-(\ref{F_g}) and is thus feasible in NOCP, with the objective value being $R^{*}(T')$. By (\ref{F_a}), $R^{*}(T')=R(\bm{x}',T')\leq R'$ holds.
\end{proof}

\noindent \textbf{Lemma 3.} $R^{*}(T)$ is monotonically decreasing in $T$.
\begin{proof}
Suppose $T_1<T_2$. Denote by $\bm{x}_1$ and $\bm{x}_2$ the optimal solutions of $R^*(T_1)$ and $R^*(T_2)$, respectively. By the proof in Appendix, $R(\bm{x}_1,T_1)\ge R(\bm{x}_1,T_2)$.
Thus, $R(\bm{x}_1,T_1)\ge R(\bm{x}_2,T_2)$.
\end{proof}

\noindent \textbf{Theorem 4.} $[\bm{x}^*,T^*]$  is optimal to DOCP if and only if $T^*=\argmin_T \{R^{*}(T)=R'\}$, where
$\bm{x}^*=\argmin\limits_{\bm{x}} R(\bm{x},T^*)|(\ref{F_f})-(\ref{F_g}) $.
\begin{proof}
For sufficiency, by Lemma~2, if $[\bm{x}^*,T^*]$  is the optimum of DOCP, $R(\bm{x}^*,T^*)\le R'$.  
 Assume strict inequality. By Lemma~1, $[\bm{x}^{*},T^{*}]$ is not the optimum. Hence $[\bm{x}^{*},T^{*}]$ satisfies $T^*=\argmin_T \{R^{*}(T)=R'\}$.
 We then prove the sufficiency. If $T^*=\argmin_T \{R^{*}(T)=R'\}$, by Lemma~2, $[\bm{x}^*,T^*]$ is feasible to DOCP. As $R^*(T)$ is monotonically decreasing in $T$, one cannot decrease $T^{*}$, otherwise the constraint (\ref{F_a}) would be violated.
 Thus, $[\bm{x}^*,T^*]$ is the optimum.
\end{proof}

By Theorem 4, $[\bm{x}^*,T^*]$ can be derived via finding $T^*=\argmin_T \{{R^{*}(T)=R'}\}$.

\vspace*{-1em}
\section{Algorithm Design and Analysis}
In general it is difficult to solve formulation (2) due to its mixed-integer and non-linear elements.
The proof of the hardness
of this problem is based on a reduction from 3-SAT problem (similar to the formal proof in \cite{TDeng2017CostJ}). In our paper, we
omit the proof due to space limitation.
To address this challenge, we propose an approximative NOCP (ANOCP) approach.
Define
$
R_\text{lb}(\bm{x},T) \triangleq \frac{1}{U}\sum_{i\in \mathcal{U}}\sum_{f\in \mathcal{F}} P_{fi}[ \frac{\max(S^f_\text{rec}-\mathbb{E}(S_{fi}),0)}{S^f_\text{rec}} ]$.
Comparing $R(\bm{x},T) $ to $R_\text{lb}(\bm{x},T) $, the difference is that the former expression has the item $\mathbb{E}[\max(S^f_\text{rec}-S_{fi},0)] $ which is in a non-linear form. The latter has the item $\max[S^f_\text{rec}-\mathbb{E}(S_{fi}),0] $ (or equivalently $\max[\mathbb{E}(S^f_\text{rec}-S_{fi}),0]  $) which can be converted to a linear form.
By mathematical analysis, one can prove that $R_\text{lb}(\bm{x},T)\le R(\bm{x},T)$. More specifically, if $S^f_\text{rec}>\mathbb{E}(S_{fi}) $, $\max(S^f_\text{rec}-\mathbb{E}(S_{fi}),0) \le \mathbb{E}( \max(S^f_\text{rec}-S_{fi},0)) $. Thus, $R_\text{lb}(\bm{x},T)\le R(\bm{x},T)$. If $S^f_\text{rec}\le \mathbb{E}(S_{fi}) $, $\max(S^f_\text{rec}-\mathbb{E}(S_{fi}),0)=0$. By considering any arbitrary number of statistics experiments, for the possible outcome one can show that $\max(S^f_\text{rec}-\mathbb{E}(S_{fi}),0) \le \mathbb{E}( \max(S^f_\text{rec}-S_{fi},0))$ also holds.
The details are provided in \cite{TDeng2017CostJ}.

Using $R_\text{lb}(\bm{x},T)$, for fixed $T$, define an ANOCP problem
\[
R_\text{lb}^*(T) \triangleq  \min\limits_{\bm{x}} R_\text{lb}(\bm{x},T)|(\ref{F_f})-(\ref{F_g}).
\]
It is interesting to examine if there can be theoretical guarantee of the difference of NOCP and ANOCP. This boils down to two objective functions and the variable $S_{fi} $ is the only factor that affects the value difference of the two. The value of $S_{fi} $ depends on the caching decisions of file $f$. The caching decision of file $f$ at user $i$ influences other files' caching decisions at this user because the cache capacity is fixed. Thus, by taking into consideration the cache capacity constraints, the values of $R(\bm{x},T) $ and $R_\text{lb}(\bm{x},T) $ are not monotonic in $x_{fi}$. Due to cache capacity, it is hard to derive a theoretical guarantee on the difference of the two. In spite of the above difficulty, in the performance evaluation section, we numerically evaluate the difference of the two expressions.

To obtain the global optimal solution of ANOCP, we introduce binary variable $y^k_{fi}$ that is one if and only if user $i$
caches $k$ segments of file $f$.
By definition, if $x_{fi}=k$, then $y^k_{fi}=1$. For example, if $x_{fi}=2$, then $y^2_{fi}=1$ and $y^k_{fi}=0$ for
$k\not=2$.
Thus, $x_{fi}=\sum_{k=0}^{S^f_\text{rec}}ky^k_{fi}$.
Define $e^k_{fij}\triangleq \mathbb{E}(\min(BM_{ij},k))$.
For any $x_{fj}$, $\mathbb{E}(\min(BM_{ij},x_{fj}))=\sum_{k=0}^{S^f_\text{rec}} e^k_{fij}y^k_{fj}$.
As a result, ANOCP can be reformulated mathematically as shown in (3), where $N_{fi}=S^f_\text{rec}-\sum_{j\in \mathcal{U},j\neq i} \sum_{k=0}^{S^f_\text{rec}}
(e^k_{fij}y^k_{fj})-\sum_{k=0}^{S^f_\text{rec}} (ky^k_{fi})$.
\begin{figure}[!h]
\vskip -20pt
\begin{subequations}
\begin{alignat}{2}
\quad &
\min\limits_{
\bm{y}}\quad  \frac{1}{U}\underset{i\in \mathcal{U}}\sum \underset{f\in \mathcal{F}}\sum
P_{fi}\frac{N'_{fi}}{S^f_\text{rec}} \\
\text{s.t}. \quad
\label{F_b}
&N'_{fi} \ge N_{fi},~ i\in \mathcal{U}, ~ f\in \mathcal{F}\\
\label{F_c}
&N'_{fi} \ge 0, ~ i\in \mathcal{U},~ f\in \mathcal{F}\\
\label{F_e}
& \sum_{k=0}^{S^f_\text{rec}}y^k_{fi}=1,~i\in \mathcal{U}, ~ f\in \mathcal{F}\\
& \underset{f\in \mathcal{F}}\sum \sum_{k=0}^{S^f_\text{rec}}ky^k_{fi}\le C_i, ~i\in \mathcal{U}\\
& \underset{i\in \mathcal{U}}\sum \sum_{k=0}^{S^f_\text{rec}} ky^k_{fi}\le S^f_\text{max}, ~f\in \mathcal{F}\\
& y^k_{fi} \in \{0,1 \},~i\in \mathcal{U}, ~ f\in \mathcal{F},~k\in [0,S^f_\text{rec}]
\end{alignat}
\label{eq:mine}
\vskip -30pt
\end{subequations}
\end{figure}
The objective function and constraints in (3) are linear with respect to $\bm{y}$. Thus, the global optimal solution,
denoted by $\bm{y}^*$,
can be obtained via using an off-the-shelf
integer programming
algorithm. Then, $\bm{y}^*$ can be straightforwardly converted into $\bm{x}$, referred to as $\bm{x}_\text{lb}$.

\noindent \textbf{Lemma 5.} For any $T'$, denote by $\bm{x}'$ and $\bm{x}_\text{lb}'$ the optimal solutions of $R^*(T')$ and $R_\text{lb}^*(T')$, respectively.
We have $R(\bm{x}_\text{lb}',T')\ge R(\bm{x}',T')\ge R_\text{lb}(\bm{x}'_\text{lb},T')$.
\begin{proof}
As $\bm{x}'$ and $\bm{x}_\text{lb}'$ are the optimal solutions of $R^*(T')$ and $R_\text{lb}^*(T')$, respectively, $R(\bm{x}_\text{lb}',T')\ge R(\bm{x}',T')$ and $R_\text{lb}(\bm{x}',T')\ge R_\text{lb}(\bm{x}_\text{lb}',T')$ for two minimization problems. Additionally, $R(\bm{x}',T')\ge R_\text{lb}(\bm{x}',T')$. Therefore, $R(\bm{x}_\text{lb}',T')\ge R(\bm{x}',T')\ge R_\text{lb}(\bm{x}_\text{lb}',T')$.
\end{proof}

\noindent \textbf{Lemma 6.} $R^{*}_\text{lb}(T)$ is monotonically decreasing in $T$.
\begin{proof}
Suppose $T_1<T_2$. Denote by $\bm{x}^1_\text{lb}$ and $\bm{x}^2_\text{lb}$ the optimal solutions of $R^*_\text{lb}(T_1)$ and $R^*_\text{lb}(T_2)$, respectively. For any $\bm{x}$, $R_\text{lb}(\bm{x},T)$ is a monotone decreasing function in $T$ (the proof is omitted due to space limitation). For example, if $\bm{x}=\bm{x}^1_\text{lb}$,
$R_\text{lb}(\bm{x}^1_\text{lb},T_1)\ge R_\text{lb}(\bm{x}^1_\text{lb},T_2)$. In addition, $R_\text{lb}(\bm{x}^1_\text{lb},T_2)\ge R_\text{lb}(\bm{x}^2_\text{lb},T_2)$. Thus, $R_\text{lb}(\bm{x}^1_\text{lb},T_1)\ge R_\text{lb}(\bm{x}^2_\text{lb},T_2)$.
\end{proof}

By Lemma 6, a bisection algorithm can be used to find $T_\text{lb}^*=\argmin_T \{R^{*}_\text{lb}(T)=R'\}$, and
$\bm{x}_\text{lb}^* = \argmin_{\bm{x}} R_\text{lb}(\bm{x},T_\text{lb}^*)|(\ref{F_f})-(\ref{F_g})$.
The algorithmic flow is presented in Algorithm 1.
If $R^*_\text{lb}(T_\text{max})>R'$, then $R^*(T_\text{max})>R'$. Thus, for any $T<T_\text{max}$, $R^*(T)>R'$ as
$R^*(T)$ is monotonically decreasing in $T$. That is, for any $T$, constraint (\ref{F_a}) cannot be satisfied, resulting in infeasibility.
\vspace*{-0.3em}
 \begin{algorithm}
\caption{The bisection algorithm}
\label{alg1}
\begin{algorithmic}[1]
\REQUIRE $T_\text{min}$, $T_\text{max}$, and $\epsilon >0$.
\ENSURE $T_\text{lb}^*$ and $\bm{x}_\text{lb}^*$
\WHILE {$T_\text{max}-T_\text{min}>\epsilon$}
\STATE $T_\text{lb}^*\leftarrow (T_\text{max}+T_\text{min})/2$
\IF {$(R^*_\text{lb}(T_\text{lb}^*)-R')(R'-R^*_\text{lb}(T_\text{min}))<0$}
\STATE $T_\text{min}\leftarrow T_\text{lb}^*$
\ELSE
\STATE $T_\text{max}\leftarrow T_\text{lb}^*$
\ENDIF
\ENDWHILE
\STATE $\bm{x}_\text{lb}^* \leftarrow \argmin_{\bm{x}} R_\text{lb}(\bm{x},T_\text{lb}^*)|(\ref{F_f})-(\ref{F_g})$
\end{algorithmic}                                                                                                                                                \end{algorithm}
\vspace*{-1.3em}

\noindent \textbf{Theorem 7.} $T^*_\text{lb}$ is a lower bound of $T^*$, i.e., $T^*_\text{lb}\le T^*$.
\begin{proof}
The result follows from
$R^*(T^*_\text{lb})\ge R^*_\text{lb}(T^*_\text{lb})$ (Lemma~5),
$R^*_\text{lb}(T^*_\text{lb})=R^*(T^*)=R'$, and
Lemma~3.
\end{proof}
 By Lemma 5, $R(\bm{x}^*_\text{lb},T^*_\text{lb}) \ge R'$, manifesting that $[\bm{x}^*_\text{lb},T^*_\text{lb}]$
 is not a feasible solution of DOCP. However, by Theorem 7, it can serve the purpose of performance benchmarking of any sub-optimal algorithm, such as Algorithm 2, because the gap to $T^*$
is less than the gap to $T^*_\text{lb}$.

Although off-the-shelf integer linear programming
algorithms can obtain solutions for up to medium-size scenarios, the
computation complexity does not generally scale. Because of this, we
propose a relaxation-rounding approach that uses continuous variables
via relaxing the integer requirement of $y^k_{fi}$, resulting in a
linear programming problem, which is polynomial-time solvable (e.g.,
by the ellipsoid method \cite{DBertsimas1997}). By the relaxation,
$y^k_{fi}\in [0,1]$ and
$\sum_{k=0}^{S^f_\text{rec}}y^k_{fi}=1$. Denote the optimum
of the relaxed problem by $\bm{y}_\text{c}$. We apply rounding to
$\bm{y}_\text{c}$.
Specifically, for any pair of $f$ and $i$, exactly one of
the variables $y^{k}_{fi}$ ($k=0,1,\dots,S^f_\text{rec}$) is rounded
to $1$. For each variable, we use its current value, which is between
$0$ and $1$, as its probability of being selected. The other variables
are set to be zeros. The corresponding integer solution is then easily
derived. By using this approach to solve $R^*_\text{lb}(T)$,
Algorithm 1 has another pair of output results, denoted by
$T_\text{c}^*$ and $\bm{x}_\text{c}^*$.
\vspace*{-0.3em}
\begin{algorithm}
\caption{The ESA Algorithm}
\label{alg1}
\begin{algorithmic}[1]
\REQUIRE $T$, $\bm{x}$, $\eta$, $\epsilon>0$, and $\eta>\epsilon$.
\ENSURE $T_\text{so}$ and $\bm{x}_\text{so}$
\STATE $T_\text{so}\leftarrow T$ and $\bm{x}_\text{so} \leftarrow \bm{x}$
\WHILE {$R(\bm{x}_\text{so},T_\text{so}) > R'$ and $\eta>\epsilon$}
\STATE $T_\text{so} \leftarrow T_\text{so}+\eta$
\IF {$T_\text{so}>T_\text{max}$}
\STATE $T_\text{so} \leftarrow T_\text{so}-\eta$
\STATE $\eta \leftarrow \eta/2$
\ENDIF
\STATE $\bm{x}_\text{so} \leftarrow \argmin_{\bm{x}} R_\text{lb}(\bm{x},T_\text{so})|(\ref{F_f})-(\ref{F_g})$
\ENDWHILE
\end{algorithmic}                                                                                                                                                \end{algorithm}
\vspace*{-1.3em}

As the next step, we propose an effective search algorithm (ESA), given in Algorithm 2,
to derive a sub-optimal
solution of DOCP and the corresponding objective function value,
denoted by $\bm{x}_\text{so}$ and $T_\text{so}$, respectively.
Initially,
for the input values of $T$ and $\bm{x}$, there are two
cases. The first case is $T=T^*_\text{lb}$ and
$\bm{x}=\bm{x}^*_\text{lb}$. Another case is $T=T^*_\text{c}$ and
$\bm{x}=\bm{x}^*_\text{c}$.
 If the first case is selected as inputs, solving $\argmin_{\bm{x}} R_\text{lb}(\bm{x},T_\text{so})|(\ref{F_f})-(\ref{F_g})$ in line 7 uses the integer programming approach. Otherwise, solving $\argmin_{\bm{x}} R_\text{lb}(\bm{x},T_\text{so})|(\ref{F_f})-(\ref{F_g})$ uses the relaxation-rounding approach. The performance evaluation section will compare the performance of the two approaches.
In each iteration,
we increase $T_\text{so}$ with step length $\eta$. If $T_\text{so}> T_\text{max}$, $T_\text{so}$ is recovered to the value of last iteration and $\eta$ is reduced by a half.
As $R^*(T_\text{so}) \le R(\bm{x}_\text{so},T_\text{so})$, $R^*(T_\text{so})\le R^*(T^*)$. By Lemma 3, $T_\text{so}\ge T^*$.

\vspace*{-0.5em}
\section{Performance Evaluations}
The effectiveness of the ESA algorithm is evaluated by comparing it to the lower bound of global optimum and conventional caching
 algorithms, i.e., random caching \cite{Balaszczyszy2015Optimal} and popular-based caching \cite{Ahlehagh2014Video}.
The file request probability follows a Zipf distribution with shape parameter $\gamma_i$ for user $i$, i.e.,
$P_{fi}=f^{-\gamma_i}/\sum_{k\in \mathcal{F}} k^{-\gamma_i}$.
The number of segments to recover a file $f$, $S^f_\text{rec}$, is randomly selected in $[1,3]$, and  $S^f_\text{max}=3S^f_\text{rec}$.
We use a Gamma distribution $\Gamma(4.43,1/1088)$ \cite{RWang2016} to generate the average number of contacts per unit time for
users $i$ and $j$, $i\not=j$, $\lambda_{ij}$. In the simulations, $C_i$ and $\gamma_i$ are uniform for all $i$, namely, $C_i=C$ and $\gamma_i=0.8$.
Besides, $T_\text{min}=0$, $T_\text{max}=400$, $\eta=1$, and $\epsilon=10^{-6}$. Our data sets are available at \cite{Dataset}.

A numerical experiment is conducted to evaluate the difference of the two expressions for a small scenario in which the number of users is three and the number of files is eight. The histogram in Fig. 1 shows the value difference of the two for all possible caching decisions of all the users and files.
It can be observed that for almost half of the possible caching
decisions the value difference is 0, and the maximum difference is
less than $8\%$. This observation manifests that the approximation is
satisfactory, and further supports problem-solving using the
approximation, in particular in view of that the approximation
overcomes the difficulty of non-linearity.

Fig. \ref{Cchanged} shows the impact of $C$ on delay. In
this figure, we use the term ILP to refer to the integer programming
based solution. We use RRA to denote the relaxation-rounding based
solution. The values of lower bound represent the results of
$T_\text{lb}^*$.  As expected, the delay decreases with
respect to $C$. The ESA algorithm surpasses the two
conventional caching algorithms. This is because the latter algorithms
do not exploit user mobility.
To further validate the effectiveness of the proposed algorithm, we consider a much larger scenario in which $N_u=30$, $N_f=1500$, $R'=0.75$, and the other parameters are kept the same as for Fig. \ref{Cchanged}.
By increasing $C$ from $5$ to $8$, for the ILP, the ranges of improvement are $[25.0\%,29.8\%]$ and $[67.3\%,84.4\%]$ over the popular and random caching algorithms, respectively. For the RRA, the ranges of improvement are $[11.5\%,14.9\%]$ and $[61.4\%,81.1\%]$ over the popular and random caching algorithms, respectively.
These improvements are larger than those for the small scenario in Fig. \ref{Cchanged}. Hence, the algorithm is suitable for large-scale optimization.
In addition, there is another observation that ILP achieves better performance than RRA. This is because the former pays the price of higher complexity due to the use of integer programming. In contrast, the latter is a polynomial time, which can be regarded as a tradeoff between complexity and accuracy.

\begin{figure}
\centering
\includegraphics[scale=0.32]{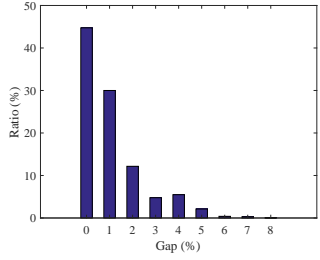}
\begin{center}
\vspace*{-1.3em}
\caption{Gaps of all possible caching decisions for a small scenario.}
\vspace*{-3em}
\label{Gap}
\end{center}
\end{figure}

\vspace{-2mm}
\section{Conclusions}
This letter has modelled a DOCP problem considering the impact of
mobility.  A lower bound of global optimum of DOCP has been
derived. For problem-solving, the ESA algorithm has been developed.
The ESA
algorithm leads to significant improvement over conventional caching
algorithms.
\begin{figure}
\centering
\includegraphics[scale=0.32]{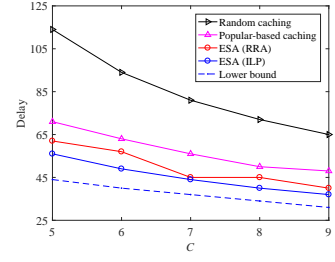}
\begin{center}
\vspace*{-1.3em}
\caption{Impact of $C$ when $B=2$, $R'=0.7$, $N_f=150$, and $N_u=20$.}
\vspace*{-2.7em}
\label{Cchanged}
\end{center}
\end{figure}

\vspace*{-2.5em}
\appendix
Suppose that $T_1<T_2$. For any $\bm{x}$, $R(\bm{x},T_2)-R(\bm{x},T_1)=\frac{1}{U}\sum_{i\in \mathcal{U}}\sum_{f\in \mathcal{F}} \frac{P_{fi}}{S^f_\text{rec}}(g(T_2)-g(T_1) )$, where $g(T)=\mathbb{E}[\max(S^f_\text{rec}-S_{fi},0)]$.
Define $\Delta g \triangleq g(T_2)-g(T_1)=\sum_{b=0}^{S^f_\text{rec}-1}(S^f_\text{rec}-b)[\text{Pr}(S^2_{fi}=b)-\text{Pr}(S^1_{fi}=b)]$.
We use mathematical induction to prove $\Delta g\le 0$. When $S^f_\text{rec}=1$, $\Delta g=\text{Pr}(S^2_{fi}=0)-\text{Pr}(S^1_{fi}=0)=\prod_{j\in \mathcal{U},j\neq i} e^{-\lambda_{ij}T_2}-\prod_{j\in \mathcal{U},j\neq i} e^{-\lambda_{ij}T_1}<0$.
Now, assume that $\Delta g \le 0$ holds for $S^f_\text{rec} = k$.
Namely, $\Delta g=\sum_{b=0}^{k-1}(k-b)[\text{Pr}(S^2_{fi}=b)-\text{Pr}(S^1_{fi}=b)] \le 0$.
When $S^f_\text{rec}=k+1$, $\Delta g=\sum_{b=0}^{k}(k+1-b)[\text{Pr}(S^2_{fi}=b)-\text{Pr}(S^1_{fi}=b)]=\sum_{b=0}^{k}[\text{Pr}(S^2_{fi}=b)-\text{Pr}(S^1_{fi}=b)]
+\sum_{b=0}^{k-1}(k-b)[\text{Pr}(S^2_{fi}=b)-\text{Pr}(S^1_{fi}=b)]$. Under the same conditions, the number of collected segments in $T_2$ is no less than that in $T_1$. Hence, $\sum_{b=k+1}^{S_\text{max}^f}\text{Pr}(S^2_{fi}=b)\ge \sum_{b=k+1}^{S_\text{max}^f}\text{Pr}(S^1_{fi}=b)$. That is, $1-\sum_{b=0}^{k}\text{Pr}(S^2_{fi}=b)\ge 1-\sum_{b=0}^{k}\text{Pr}(S^1_{fi}=b)$. Therefore, $\sum_{b=0}^{k}\text{Pr}(S^2_{fi}=b)\le \sum_{b=0}^{k}\text{Pr}(S^1_{fi}=b)$, leading to $\Delta g \le 0$.
 Based on the above proof, we can conclude $R(\bm{x},T_2)\le R(\bm{x},T_1)$ when $T_2>T_1$, proving that $R(\bm{x},T)$ is a monotone decreasing function with respect to $T$.
Moreover, it is obvious that the function is continuous.

\end{document}